\documentclass[12pt]{article}
\usepackage{epsf,epsfig,graphicx}
\usepackage{cite}
\usepackage[dvips]{color}
\usepackage[normalem]{ulem}%

\definecolor{darkgreen}{cmyk}{1,0,1,0.4}

\newcommand{\beq}{\begin{equation}}
\newcommand{\eeq}{\end{equation}}
\newcommand{\ben}{\begin{eqnarray}\displaystyle}
\newcommand{\een}{\end{eqnarray}}
\newcommand{\bea}[1]{\begin{eqnarray}\label{#1} }
\newcommand{\eea}{\end{eqnarray}}
\newcommand{\refb}[1]{(\ref{#1})}

\newcommand{\sectiono}[1]{\section{#1}\setcounter{equation}{0}}

\def\boxempty{{\,\lower0.9pt\vbox{\hrule \hbox{\vrule height 0.25 cm
\hskip 0.25 cm \vrule height 0.25 cm}\hrule}\,}}
\def\one{{\hbox{ 1\kern-.8mm l}}}
\def\zero{{\hbox{ 0\kern-1.5mm 0}}}

\def\gtap{\raisebox{-.4ex}{\rlap{$\sim$}} \raisebox{.4ex}{$>$}} 

\begin{document}
\begin{titlepage}
\thispagestyle{empty}

\title{
{\Huge\bf Standard Cosmology Delayed
}}

\bigskip\bigskip

\author{
{\large\bf Debajyoti Choudhury}\\
{\large\it Department of Physics and Astrophysics, University of Delhi}\\
{\large\it Delhi 110007, India}\\
{\tt debajyoti.choudhury@gmail.com}\\
{}\\
{\large\bf Debashis Ghoshal}\\
{\large\it School of Physical Sciences, Jawaharlal Nehru University}\\
{\large\it New Delhi 110067, India}\\
{\tt dghoshal@mail.jnu.ac.in}\\
{}\\
{\large\bf Anjan Ananda Sen}\\
{\large\it Centre for Theoretical Physics, Jamia Millia Islamia}\\
{\large\it New Delhi 110025, India}\\
{\tt anjan.ctp@jmi.ac.in}\\
}

\bigskip\bigskip

\date{
\begin{quote}
\bigskip\bigskip \centerline{\bf Abstract} 
{\small 
  The introduction of a delay in the Friedmann equation of cosmological
  evolution is shown to result in the very early universe undergoing
  the necessary accelerated expansion
  in the usual radiation (or matter) dominated phase.
  Occurring even without a violation of the strong energy condition,
  this expansion slows down naturally to go over to the
  decelerated phase, namely the standard Hubble expansion.
  This may obviate the need for a scalar field driven inflationary epoch.
}
\end{quote}
}



\end{titlepage}
\maketitle\vfill \eject


\section{Introduction}\label{introd}%
The Standard Model of Cosmology, based on assumptions of large scale 
statistical homogeneity and isotropy,  is an elegantly simple and powerful
theory. Supplemented with the Standard Model of Particle Physics---both 
of which are continuum field theories based on local interactions---it
explains the evolution of our expanding universe almost up to the point 
of the initial big bang singularity.  However, despite this success, there 
are issues that remain unresolved---indeed, for various reasons it seems 
unlikely that these can be the fundamental theories.

As our understanding of local quantum field theories has deepened, it
has become increasingly clear that any such theory is only an effective
description valid at an appropriate scale. The existence of Planck units 
as scales of length and time (deduced purely on dimensional grounds) 
seems to point to a limit to which one may hope to push this formalism. 
String theory (from which local field theories emerge in the low energy 
limit) is an alternative formalism that incorporates non-local interactions 
at Planck scale\cite{string_nonloc}. Similarly, loop quantum gravity or 
spin foam too have an inherent non-locality associated with 
them\cite{Markopoulou:2007ha,Ng:2008pi}.
In a relativistically invariant theory, nonlocality also implies interactions 
smeared in time and delayed reaction, in particular.

Delayed reaction in dynamical systems occupy a place of central
importance in many areas of science. In particular, many biological
systems are governed by delay differential equations (DDE)\cite{Erneux}.
Examples are models of population growth\cite{MurrayMB}, where they 
first made their appearance. A characteristic feature of DDE is the way 
the derivative at any instant of time depends on the function at earlier 
times.  This translates to solutions radically different from those obtained 
in the zero-delay case. To wit, a first order linear DDE with real coefficients
may admit oscillatory solutions. Delayed dynamics in the decay of unstable 
branes of string theory  have been studied\cite{Hellerman:2008wp} and 
compared to biological systems\cite{Barnaby:2010kx}.

In this article, we regard cosmological evolution as a dynamical 
system\cite{EllisWW, Coley}. We show that the introduction of a delay 
in the Friedmann equation ameliorates some of the shortcomings of 
the standard cosmological model. For a rather general class of initial 
data in, say, the radiation dominated epoch, the early universe is found to 
undergo a brief phase of accelerated expansion. Further, this slows 
down naturally to a decelerated expansion and asymptotes to standard 
FRW evolution. Thus, a delayed reaction within standard cosmology 
seems to obviate the need for an inflationary epoch driven by a scalar 
field as demanded in the standard paradigm and, simultaneously, 
solves the associated graceful exit problem. What is more, the initial 
accelerated expansion is obtained without a violation the strong energy 
condition.

Admittedly, our model is phenomenological in its spirit. We shall not 
attempt to obtain the delayed dynamical equation from first principles. 
We recognize that this important issue needs to be addressed---for an 
approach see \cite{Atiyah:2010qs}. And while a phenomenological model 
of  non-local gravity was shown to give accelerated 
expansion\cite{DeserWood}, models of this  type are demonstrated to be 
equivalent to multi-scalar-tensor theories\cite{Koivisto:2008}. Non-local 
effects motivated by string theory have previously been studied in the 
cosmological context, see {\em e.g.}, \cite{JoukoNLCos,bkmNLCos}.


\section{The Friedmann Equation}\label{std-cosmo}%
Let us briefly recall a few essential facts of the standard FRW cosmology, 
described by the scale factor $a(t)$ that determines physical distances at 
time $t$. The dynamics of the universe at large scales is governed by the 
Friedmann equation
\begin{equation}\label{Fried_one}
\left({\frac{\dot{a}(t)}{a(t)}}\right)^2 = \frac{1}{3}\rho(t),
\end{equation}
where $\rho(t)$ denotes the total energy density and following standard 
practice, we have used natural units, namely $8\pi G = c = 1$. Homogeneity 
and isotropy at large scales dictate that $a$ and $\rho$ depend only on time. 
Assuming that the universe expands adiabatically, the first law of 
thermodynamics gives 
\begin{equation}\label{thermo_one}
{a^3}\,{\Delta\rho} + (\rho +p)\,\Delta(a^3) = 0.
\end{equation}
Combined with Eq.\refb{Fried_one}, this results in an equation for the 
Hubble expansion parameter $H(t) \equiv \dot{a} / a$:
\begin{equation}\label{HandP}
2\dot{H}(t) + 3H^2(t) = -p,
\end{equation} 
a relation that can be derived as well from the Einstein-Hilbert action of GTR. 
One may rewrite it as 
\begin{equation}\label{accel}
\frac{\ddot{a}(t)}{a(t)} = - \frac{1}{6} \left[\rho(t) + 3 p(t)\right],
\end{equation}
to see that, for an accelerated expansion to occur, $\rho(t)+3p(t)$
must be negative, {\em i.e.}, the strong energy condition (SEC) needs
to be violated. Clearly, this is not the case with usual matter or
radiation. One way to achieve this is the inflationary
paradigm\cite{Guth,Linde,Inflatn}, in which one considers a scalar
field that rolls on a sufficiently flat potential. Such models are
considered to be the most simple, and variants of this theme are
widely used to model an accelerating universe.

We note that of the three equations \refb{Fried_one}, \refb{thermo_one} and
\refb{HandP} (or equivalently, \refb{accel}), only two are independent. For
example, the third follows when we take a time derivative of the first and use
Eq.\refb{thermo_one}; and this is the approach we shall take in the following.  

\section{Delayed Friedmann Equation}\label{del-frw}%
We propose a modification of the central dynamical equation \refb{Fried_one} 
by introducing a constant delay $\tau$. Thus, we postulate that any change of 
the matter content in the universe has a delayed effect on the evolution of the 
metric and hence described by the {\em delayed Friedmann equation} 
\begin{equation}\label{Fried_del}
\left({\frac{\dot{a}(t)}{a(t)}}\right)^2 = \frac{1}{3}\rho(t-\tau).
\end{equation}
This may modify the dynamics of the universe, and we shall argue that that 
has the potential to provide a solution to some of the problems of the Standard
Cosmology. 

The standard lore may, however, prejudice one against such a nonlocal
deformation. Nevertheless, the fact is that any direct knowledge of
dynamics goes down, at best, to time scales of 
$ {\cal O}(10^{-27}\mathrm{ s})$. Thus, for $\tau$ much smaller 
than this, say, comparable to Planck time $t_{\mathrm{Pl}}\sim 
{\cal O}(10^{-43}\mathrm{ s})$, changes such as that in
Eq.\refb{Fried_del} are entirely consistent with observations.
Moreover, the microscopic
theories\cite{string_nonloc,Markopoulou:2007ha,Ng:2008pi} presently in
vogue incorporate nonlocality in an essential way. Besides, gravity
may well be an emergent phenomenon\cite{emerg_grav,Jacobson:1995ab}
(see also \cite{Visser,Padmanabhan:2009vy}), in which case the
equations related to gravity are not bound to be local. 

It should be noted, though, that while we have listed possible 
quantum (gravity) effects as the source of this delay, the rest of 
our analysis is purely classical. This, indeed, is in the spirit of
effective theories.\footnote{Analogous is the case of conventional
inflatonary theories wherein the inflaton potential is related to
supergravity or even string theoretic constructions.} And, as we 
shall show later, the numerical value of $\tau$ could be commensurate 
with energy scales far lower than $M_{\mathrm{Pl}}$ ({\em i.e.}, 
$\tau$ larger than $t_{\mathrm{Pl}}$), thereby allowing us to neglect 
quantum gravity effects, at least in the first approximation. As the 
discerning reader would recognize, this approximation is intrinsic 
to all current theories of inflation.

The dynamics of the universe, in this scheme, is thus governed by 
Eqs.\refb{thermo_one} and \refb{Fried_del}, which, in turn, imply
\begin{equation}
2\dot{H}(t) + 3H^2(t) = \rho(t-\tau) - \frac{H(t-\tau)}{H(t)}\left(\rho(t-\tau)
+ p(t-\tau)\right),\label{HandP_del}
\end{equation}
in place of Eq.\refb{HandP}.
Let us reiterate that the proposed delayed dynamics is {\em ad hoc}
and not derived from a `fundamental theory'. It might be argued that 
modifications may be made in various other ways. While this is certainly 
true, it turns out that many of the theoretically desirable changes lead 
to qualitatively similar behaviour. Hence, rather than discussing each 
alternative separately, we restrict ourselves to the one proposed above. 
We shall comment briefly on the other possibilities at the end.

Before we proceed to explore the consequences of the delay, let us emphasise
that the smallness of $\tau$ will ensure that there is virtually no change in the
late time evolution of the universe, or in the dynamics of heavenly bodies that
evolve at macroscopic scales. In other words, there would be no discernible
consequences of this modification on astrophysical scales.

\sectiono{Delayed Dynamics \&\ Early Accelerated Expansion}\label{del-dyn}%
In this section, we shall find a solution to the set of equations \refb{thermo_one},
\refb{Fried_del} and \refb{HandP_del} and discuss its properties.  
For the sake of simplicity, let us assume an equation of state of the form 
$p = w\rho$ (where $w$ is a constant) for matter or radiation that permeates 
the universe at the earliest epoch. The value of $w$ is $1/3, 0$ and $1$, 
respectively, for radiation, non-relativistic dust and stiff fluid. While it is true 
that the early universe had several different components of matter, the one 
with the largest $w$ would have dominated dynamics at early epochs. In any 
case, the inclusion of multiple fluids do not materially affect our analysis.

\begin{figure}[ht]
\vspace*{-50pt}
\centerline{
\epsfig{file=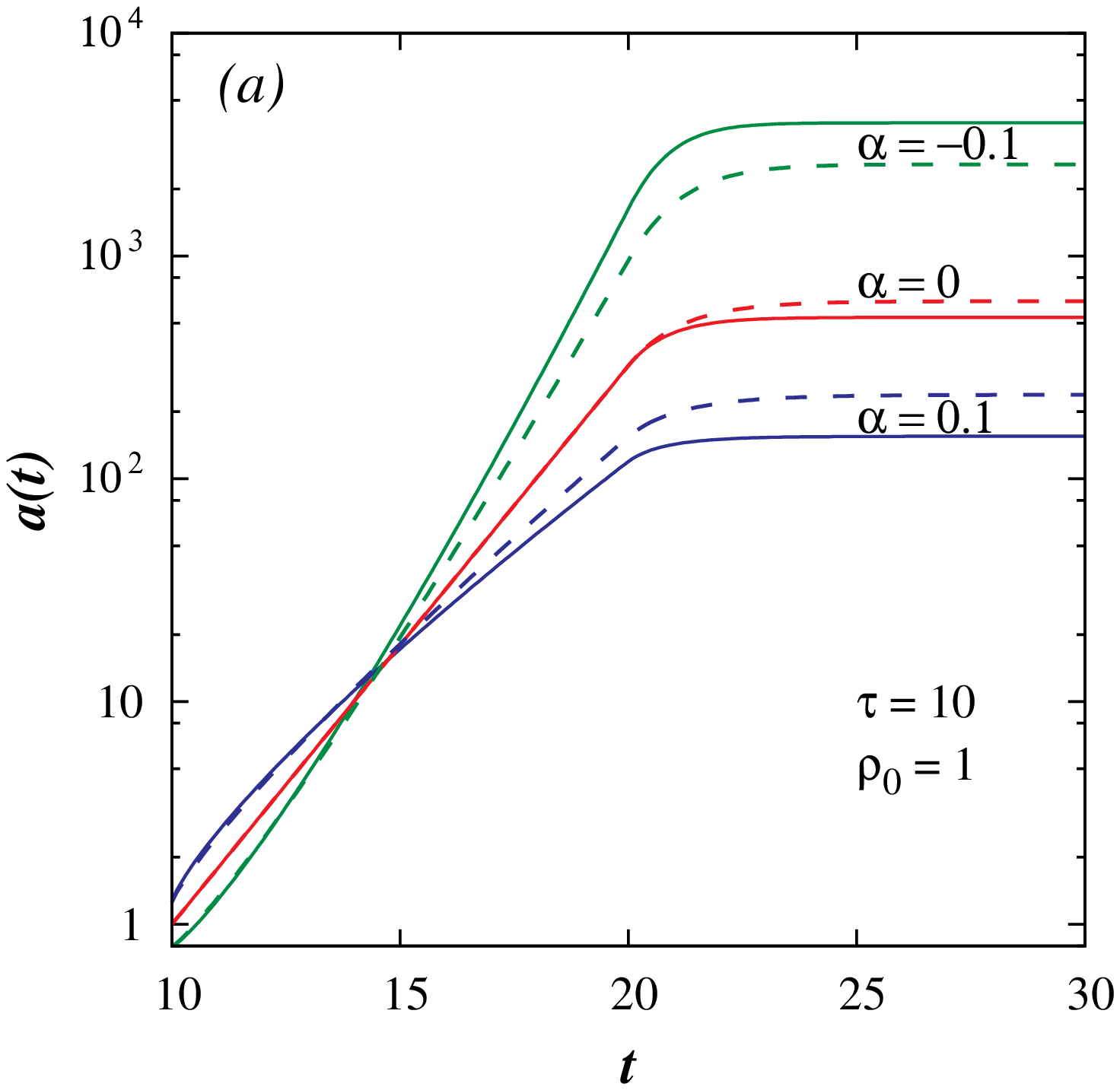,width=9cm,height=10cm,angle=0}
\hspace*{-30pt}
\epsfig{file=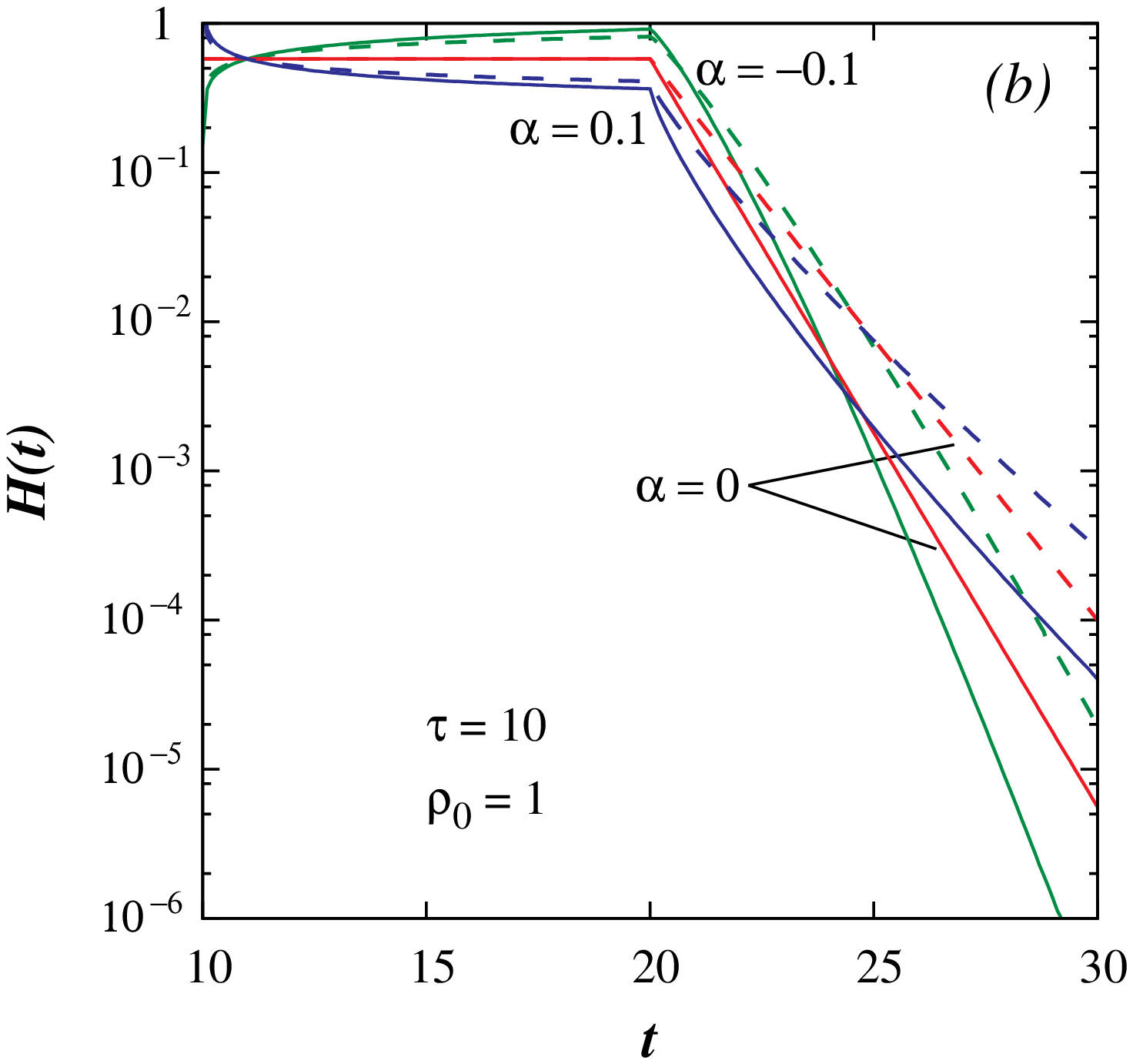,width=9cm,height=10cm,angle=0}
}
\vspace*{-70pt}
\caption{(a) The scale factor and (b) the Hubble parameter
in the first two intervals. The solid and dashed line correspond to 
$w= 1/3$ and $0$, respectively. }
\label{fig:a_of_t}
\end{figure}

The equation of state, together with the first law of thermodynamics relates 
$\rho = \rho_{0}a^{-3(1+w)}$, where $\rho_{0}$ is an arbitrary constant of 
integration. Substituting this in Eq.\refb{Fried_del}, we have
\begin{equation}
\frac{d}{dt} \ln a(t) = \sqrt{\frac{\rho_{0}}{3}} \;
 \left[ a(t-\tau)\right]^{-3 \, (1+w) / 2}.
\end{equation}
We solve this using the method of steps\cite{Erneux} starting with
the `initial condition' $a(0 \leq t < \tau) = f(t)$ where $f(t)$ is a
given function.  For definiteness, let us consider $f(t) = t^\alpha$
with a constant $\alpha$.  (Note that, as long as $\alpha \leq 1$, the
universe is actually decelerating in this epoch, with $\alpha < 0$
implying a collapse.  Such a situation may develop for a variety of
reasons, including quantum gravity effects or a prior (pre `Big Bang')
history of the universe---it is not derived from within our model.) It
now amounts to solving an ODE in subsequent intervals of size
$\tau$. Of particular interest is the solution in the first interval:
\begin{equation}\label{inflation}
a(t) = \tau^\alpha\,\exp\left(\sqrt{{\rho_0\over 3}}
 \, \frac{ (t-\tau)^{1- {3\over 2} (1+w) \, \alpha}}
         {1- {3\over 2} (1+w) \alpha} \right),\; 
\mbox{\rm for } \tau\le t < 2\tau.
\end{equation}
It is evident from the above that an accelerated expansion or {\em inflation} is 
possible for a wide choice of $\alpha$ even with normal radiation or matter. 
For later times, the solution has to be obtained numerically. We display a few 
cases in Fig.\ref{fig:a_of_t}. 

The following features of `delayed FRW dynamics' seem worth pointing out.
\begin{itemize}
   \item 
   A phase of fast growth exists for a wide range of $\alpha$. In fact, for 
   $\alpha \le 0$ (an initially static or even contracting phase), the universe
   expands exponentially or faster. A fast growth can also occur for $\alpha>0$ 
   as long as $(1+w) \, \alpha < 2/3$. (Many features actaully depend on 
   this combination of $\alpha$ and $w$.) Beyond this value, the growth of
   the scale factor is decelerated, as in standard cosmology. 
   \item 
   Throughout the phase of accelerated expansion, the strong energy
   condition holds good, in direct contrast to received wisdom, and, unlike 
   all models of accelerated growth known so far. Indeed, the usual condition 
   for acceleration, namely $w < -1/3$ is now replaced by 
   \begin{equation}
        \frac{H(t-\tau)}{H(t)} < \frac{2}{3(1+w)} \ .
     \end{equation}
   The requirement of $H(t-\tau)/H(t)$ being less than $1/2$ for radiation (or 
   $2/3$ for nonrelativistic dust), is easy to satisfy for a large class of initial 
   conditions. This, in essence, is the most important and interesting result to
   emerge out of the proposed delayed dynamics.
  \item 
   The rate of growth is determined by the initial matter density $\rho_0$ and 
   grows with it. Again, this would seem counterintuitive, for a larger energy 
   density, instead of slowing down the expansion, actually increases it.
   \item   
   The end of inflation is denoted by the onset of deceleration. In the present 
   case, there is a subtlety. While, for $(1+w)\alpha \leq 0$, the universe has 
   an accelerated expansion during $\tau \leq t < 2 \tau$, for 
   $0 < (1+w)\alpha < 2/3$, on the other hand, the universe is decelerating 
   initially, but quickly passes onto an accelerating phase, with the onset of 
   acceleration being progressively delayed for larger values. The initial 
   decelerating phase, however, would leave virtually no signal in the sky. 
   \item
   The duration of the `inflationary' growth phase is of the same order
   as the delay $\tau$, as the universe quickly settles down to a phase of 
   decelerated growth after $t \gtap 2 \tau$ (see Fig.\ref{fig:a_of_t}{(b)}). 
   In most conventional models of inflation, `graceful exit' is often a 
   problem and a mechanism needs to be introduced to ensure that the 
   accelerated expansion stops. Remarkably, the introduction of a delay 
   in Eq.(\ref{Fried_del}) not only introduces an inflationary phase, but also 
   serves to bring the universe out of it. 
   Naively, one would have expected that the exit would occur only for 
   $t \gg \tau$, as, at such late times, the delay would be immaterial and the
   system would essentially revert to the normal power law expansion. Rather 
   the exit is {\em precocious} in the present system. Fortunately, however, 
   this does not cost us in terms of the number of e-foldings. 
\end{itemize}

The exit from inflation is accompanied by an abrupt change in $\ddot{a}(t)$ 
(see Fig.\ref{fig:a_of_t}(b)).  Such discontinuities (one such also occurs at 
$t = \tau$) in higher derivatives are endemic to DDEs with generic initial 
conditions. In the present context, this could potentially affect primordial 
density perturbations and consequent signatures in the background microwave 
spectrum as also the production of superheavy dark matter~\cite{Riotto:1998}. 
Further discussion of this, however, is beyond the scope of this work.

\begin{figure}[ht]
\begin{center}
\includegraphics[scale=0.6]{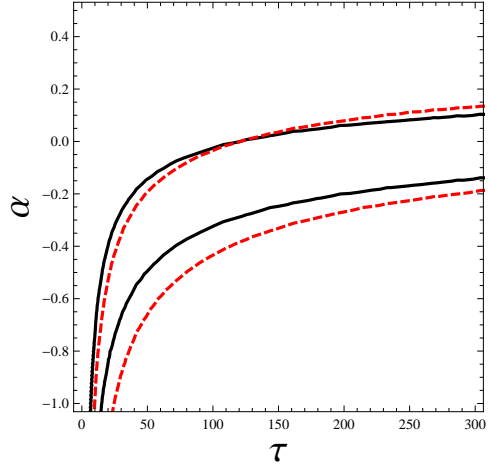}
\end{center}
\caption{The contours for the number of e-folds in the $\alpha$-$\tau$ plane. 
The solid and dashed line correspond to $w= 1/3$ and $0$, respectively. In 
each case, the upper curve is for 70 e-folds while the lower one is for 700
e-folds.} 
\label{fig:efold}
\end{figure}

It is well known that for inflation to solve the problems of standard
cosmology, we need at least about 65 e-folds ({\em i.e.}, growth in
$a(t)$ by a factor of $e^{65}$) by the end of inflation.  In
Fig.\ref{fig:efold}, we display constant-inflation contours in the
$\tau$-$\alpha$ plane for different values of $w$. Note that the
dependence on $w$ is small. In effect, the bulk of inflation occurs in
the first epoch ($\tau \leq t \leq 2 \, \tau$) where the dynamics is
governed by the product $(1+w)\alpha$. It is only in the next phase,
where the expansion slows down considerably and the `inflation' ends
that the separate dependence on $w$ appears. This was to be expected
in view of Fig.\ref{fig:a_of_t}. What is, perhaps, more interesting is
that the delay required becomes smaller as $\alpha$ becomes more
negative. In other words, the faster the universe was collapsing, the
faster it rebounds back; since a nonzero $\tau$ implies that crunching
of matter reacts on the space-time fabric only after a delay, it
stands to reason that a larger compression needs to be bottled up only
for a shorter duration before it reacts violently.

It is also quite apparent that, to achieve a phenomenologically
acceptable amount of inflation starting with a static or slowly
evolving `initial condition' ({\em i.e.}, a small $|\alpha|$), 
one would require $\tau \sim {\cal O}(10^{2}\mbox{--}10^{3})
t_{\mathrm{Pl}}$. One might interpret this as being a restatement 
of the requirement of the (usual) inflaton scale being 
$\sim M_{\mathrm{GUT}}$. The analogy, however, is not exact.
Rather, the amount of inflation can be approximated to be given 
by a certain function $F_{\mathrm{in}}(\tau, (1 + w) \, \alpha )$ 
with only subsidiary dependence on other variables. The functional 
form of $F_{\mathrm{in}}$ does depend on the exact initial condition 
$a(0 \leq t < \tau)$, but, as we have demonstrated above, 
inflation occurs for a very wide class of initial conditions. It 
is worthwhile to point out at this stage that, contrary to popular 
wisdom, it is not necessary that inflation must occur at 
$M_{\mathrm{GUT}}$ or thereabouts. For example, as shown in 
Ref.\cite{Padmanabhan:1988se}, the invocation of a Planck-size 
proper-length cut-off automatically regulates the size of the 
density perturbations to acceptable limits. While the cut-off in 
Ref.\cite{Padmanabhan:1988se} may have been introduced in an ad-hoc 
manner, its existence would be natural in theories that would lead 
to delays, such as ours. 


\section{Endnote}\label{summ}%
In this paper, we have initiated the study of an alternative mechanism 
for an accelerated expansion in the early universe. It is not driven by a scalar 
field, but rather achieved by modifying the Friedmann equations by the 
introduction of a delay, in a phenomenological fashion.
We are aware that many questions have remained unanswered and our
first effort in this direction is far from complete. While the answers to 
these would evidently require major effort, we shall touch upon some 
of these issues presently.

\begin{itemize}
\item {\em On the completeness of the theory:}
 
  Let us reiterate that although non-locality abounds in quantum theories of
  gravity\cite{string_nonloc,Markopoulou:2007ha,Ng:2008pi,Hellerman:2008wp,%
  Atiyah:2010qs,JoukoNLCos,bkmNLCos} and we have been motivated by
  its ubiquity, ours is only a phenomenological approach. Ideally, one
  should have a complete theory of gravity to study all the
  ramifications associated with the idea of inflation. It is needless to say
  that a naive modification of the Einstein equations cannot be the 
  answer. The ideas propounded in Ref.\cite{Atiyah:2010qs}
  demonstrates that delayed dynamics in a covariant framework requires
  much subtlety. Meanwhile, we may consider an analogous dynamical
  equation describing the decay of an unstable
  brane\cite{Hellerman:2008wp,Ghoshal:2011rs}.  It is conceivable that
  once such branes are coupled to gravity, similar effects, in the
  presence of an appropriate background, would lead to delayed
  equations involving gravity. Admittedly, the complete dynamical
  system would be a much more complicated one than that we have
  considered. Thus, we only report on a preliminary study intended to
  stimulate the search for such field configurations.
\item {\em On the uniqueness of the delay mechanism:}

It might rightly be claimed that the delayed dynamics we consider is not 
unique. Let us comment briefly on other possible ways of introducing 
delay in the dynamics. For example, if, in addition to \refb{Fried_del}, there
is an identical delay in the RHS of \refb{accel} (in which $\rho$ and
$p$ are taken to be related by the equation of state), the evolution
of the scale factor is exactly the same as in the standard FRW
universe. However, the evolution of matter would differ. On the other
hand, suppose we work with Eqs.\refb{Fried_del} and \refb{HandP}
(which is {\em not} delayed), the time derivative of $\rho$ turns out
to depend on $a$ and $\rho$ at a time in the future, leading to an
acausal behaviour.

It is of course possible to introduce a delay in many other ways,
however, in some cases, the properties of the solution remain
qualitatively same. For instance, Eqs.\refb{Fried_one} and
$\dot{\rho}(t) + 3(1+w) H(t-\tau)\rho(t)=0$, a variant of
Eq.\refb{thermo_one}, gives exactly the same evolution. If instead,
one considers the variant $\dot{\rho}(t) + 3(1+w) H(t)\rho(t-\tau)=0$,
the qualitative behaviour of the evolution of the scale factor remains
the same.

In a sense, therefore, we have introduced a {\em minimal} modification
in which the delay occurs in only one equation, namely the Friedmann
equation, which describes the effect of matter on the expansion of the
universe.  Eq.\refb{thermo_one} is the statement of energy
conservation; and since that is applicable in a wider context, it has
not been modified.
\item {\em On the stability of the system:}

  Higher derivative theories are usually known to suffer from an inherent 
  instability, known as the Ostrogradskian instability, which was extended 
  to infinite number of higher derivatives in, {\em e.g.}, Refs.\      
  \cite{Woodard:2000bt,Woodard:2002bx}. It might be argued, with a delay 
  being equivalent to the existence of an arbitrarily higher order derivatives, 
  that the theory would necessarily be unstable. One should, however, 
  be cautious to come to this conclusion without a detailed analysis. For 
  one, the nonlocality due to the delay by $\tau$, being equivalent to
  $e^{-\tau\partial_t}$, is through an entire function of the derivative
  (with respect to time)\cite{Woodard:2000bt}. Moreover, such systems have 
  been studied in the literature, with particular emphasis on the cosmological 
  context in, {\em e.g.}, Refs.\cite{Barnaby:2006hi,JoukoNLCos,Barnaby:2010kx} 
  (and references therein) and seem to exhibit a reasonable dynamical behaviour. 
  It is, therefore, not unreasonable to expect that once the delayed 
  Friedmann equation (perhaps in a form modified from the one we use) is 
  obtained from a fundamental Lagrangian upon the inclusion of quantum 
  corrections, the corresponding dynamical system would, generically, 
  represent a stable system.

\item {\em On the dependence on the initial data:}

Modifying finite order differential equations to delayed
  differential equations has necessitated the imposition of initial
  conditions over a {\em finite continuous segment}. It might be
  argued that this represents the introduction of an infinite number of new
  degrees of freedom.  While this criticism is valid {\em per se}, we emphasise
  that the existence of an accelerating phase is not tied to a
  particular form of the initial data $f(t)$. We have checked numerically 
  that the accelerated expansion is quite generic and occurs for a very 
  wide choice of initial data. It is only that a monomial or a single exponential
  form for $f(t)$ permits a simple closed analytical form for the scale factor 
  $a(t)$, and we have chosen to illustrate our arguments with the former. For 
  more complicated forms for $f(t)$, the entire solution has to be obtained
  numerically.  

  We do not have {\em a priori} arguments in favour of any particular choice 
  for $f(t)$, preferring to demonstrate the imprint of the delay on the early 
  universe with a simple class of initial conditions. 
  These could be a result of either a series of quantum fluctuations or the 
  result of some cataclysmic {\em pre-Big Bang}\footnote{Note that, in our 
  context, it is $t = \tau$ that defines the beginning of time as we know it.}
  events (the latter possibility appears naturally in many theories of quantum 
  gravity, including, but not limited to string theory). Indeed, as long as the 
  {\em pre-Big Bang} universe was not expanding as fast as the corresponding 
  un-delayed Friedmann equation would have wanted it to, a phase of  
  accelerated expansion phase would necessarily occur in the delayed 
  version.
\end{itemize}

Finally, to summarise, we have demonstrated the possibility that a delay
introduced in the Friedmann equation could naturally lead to an
exponential (or faster) growth phase in the very early universe. The
existence of such a phase requires neither the existence of a scalar
field (inflaton) with a flat potential nor even a violation of the
strong energy condition. While our formulation is obviously a
phenomenological one, it should essentially be considered a proof of
principle motivated by the existing theories of quantum gravity.
However, a more detailed construction based on a microscopic theory
should be sought.

Not only does the universe inflate, it also asymptotes to FRW cosmology, 
thereby eliminating the need for an exit mechanism.  The required delay is 
small (a few hundred Planck times, at most) and natural in the context of 
nonlocalities inherent in quantum gravity. It is also consistent with all 
observations. Work on subsequent reheating and generation of primordial 
fluctuations are in progress and will be reported elsewhere.
Also of interest is the potential that a significant
fraction of the primordial energy density could have existed in the
form of magnetic fields (thereby offering a possible seed for the
intergalactic magnetic field observed today), a scenario that is
difficult to accommodate in canonical inflationary
scenarios\cite{Demozzi:2009fu}. A rich tapestry of many such physical
consequences may be expected.

\bigskip

\noindent{\bf Acknowledgement:} It is a pleasure to thank S.~Mukohyama, 
T.R.\ Seshadri and especially R.~Ramaswamy for discussions.  
This work was supported by SERC, DST (India) through the grant 
DST-SR/S2/HEP-043/2009. DC thanks the IPMU and the WPI Initiative, 
MEXT, Japan while AAS thanks the IUCCA, Pune for the hospitality provided 
while part of the work was being done.

\newpage



\end{document}